\documentclass[11pt]{article}
\def\half  	{\textstyle {1 \over 2} \displaystyle}

\def\inv16  	{\textstyle {1 \over 16} \displaystyle}
\def\3ov2  	{\textstyle {3 \over 2} \displaystyle}
\def\4ov3  	{\textstyle {4 \over 3} \displaystyle}

\def\beq{\begin{equation}} \def\eeq{\end{equation}}
\def\bea{\begin{eqnarray}} \def\eea{\end{eqnarray}}

\def\cD{{\cal D}}

\def\cH{{\cal H}}
\def\cI{{\cal I}}
\def\cJ{{\cal J}}

\def\cL{{\cal L}} 	\def\bcL { {\mbox{\boldmath $\cal L$}} }

\def\cN{{\cal N}}
\def\cO{{\cal O}}

\def\cQ{{\cal Q}} 
\def\cR{{\cal R}} 	
\def\cS{{\cal S}} 	
\def\cT{{\cal T}} 	\def\bcT { {\mbox{\boldmath $\cal T$}} }

\def\cZ{{\cal Z}}

 		 	\def\be{ {\bf e}} 
 	 	
\def\bv{{\bf v}} 	\def\bw{ {\bf w}}	 	 	
 	\def\bX{{\bf X}} 	\def\bY{{\bf Y}} 	

\def\bOne { {\mbox{\boldmath $1$}} }
\def\UL{\underline{~}}

 		\def\bvarep { {\mbox{\boldmath $\varepsilon$}} }

\def\Del { {\mbox{\boldmath $\nabla$}\!} }

\def\Mink{{R$^{1,3}$}}
\def\sdprod {\,{\supset \!\!\!\!\!\!\! \times}\,}

\begin{document}

\title{Baby Steps in Quantum Ring Theory: \\
towards a background independent framework for Quantum Gravity
}

\author {Rafael A.
Araya-G\'ochez\footnote{
  	{\small Physics Department, Occidental College M21,}
	{\small 1600 Campus Rd., Los Angeles, CA 90041} 
  	}
}

\date{}
\maketitle					

\begin{abstract} 	
	We investigate gravity as a gauge theory in the language of fiber bundles with tools from algebraic geometry.
 Compelled by the construction of the Eilenberg-MacLane classifying space {\it via} Fox derivations in an integral group ring,
 the origin of locally curved spacetime as a an abelian group extension is sought algebraically by means of a suitable derivation rule
 that conforms with the algebraic content of Einstein's equation.  Accordingly, a peculiar flavor of background independence is advanced;
 curved spacetime should be thought of as an abelian proxy for the Lorentz group, causal and spin throughout. 
 Thus, diffeomorphism invariance amounts to ``choosing" an appropriate algebro-geometric--differential tool as to make algebraic sense
 of Einstein's equation or of a sensible evolution of such an equation.  In this regard, it is suggested that a left-$\alpha$ derivation rule 
 may facilitate the construction. 
 On the other hand,
 we are led to interpret ``spacetime curvature" in the broad sense of the exterior covariant derivative of a {\it Cartan} connection 
 which includes both the torsion of the vierbein as well as the standard curvature of the spin connection.
 Making manifest use of the index structure in the standard Hilbert action--in particular, the distinction between internal and external indices--inclusion
 of the torsion 2-form is found to be inconsistent with contraction of the spacetime curvature as a scalar density.
 Aiming to erect a geometrically justified gravitational action that incorporates torsion,
 we lay out several fundamental geometric, differential and algebraic building blocks	
 while highlighting descent relations among the different pieces. 
 The groundwork is thus laid out for the search of the sought after action functional.  
\end{abstract}    

\pagebreak

\tableofcontents
\pagebreak 		

\section{Introduction}
 \label{sec:Intro}

\begin{quote}
{\footnotesize {\it
La voce narr\`{o}	\\ 
all'ultimo che  	\\
sul mondo rest\`{o},	\\ 
la vera realt\`{a}	\\ 

E poi comand\`{o}	\\ 
di andare tra i suoi 	\\
a dire la verit\`{a}	\\ 
e il gioco inizi\`{o}.	\\ 
}
\begin{flushright}
Ill Balleto di Bronzo: YS (1973)
\end{flushright} }
\end{quote}

 	{\Large \cQ}uantum field theory--the fundamental bastion of modern theoretical physics--relies on the notion of number fields 
 {\it indexed} by differential labels to represent particles and interactions through gauge potentials.   
 For instance, the electromagnetic vector potential is succinctly written as an integrand (i.e., a 1-form in the language of exterior differentials): 
 $A(x) = A_\mu(x^\nu) $d$x^\mu$, naturally embedded in the spacetime as non-dynamical degree of freedom of the theory.
 Notably, this construction is intrinsically differential; 
 the vector potential is simply defined by its differential over the spacetime {\it modulo} a physically irrelevant constant. 
 Conversely, the notion of ``position" in the spacetime constitutes a more subtle issue requiring abstract mathematical machinery in its 
definition\footnote{See, e.g., pg. 49 of {\it ``Gauge Fields, Knots and Gravity"}\cite{Baez&Muni94}.
}.  Although positions in the spacetime: $x \in M$, are tacitly represented by the ``pullback" of coordinate functions, $\phi^* x^\nu$, 
 on local charts $U \subset M$ of the spacetime modeled as a {\it manifold}: $\phi: U \rightarrow {\rm R}^4$ (i.e., locally {\it homeomorphic} to Euclidean space),
 the finer details of such a seemingly innocuous representation are central to the particular flavor of background independence we advocate in this note.

 	The distinction in indexing by differential labels as opposed to the labeling by coordinate functions constitutes a key fundamental difference that 
 separates quantum mechanics from quantum field theory: the former relies on single particle observables, position and momentum, as canonically conjugate 
 operators while the latter leans on the notion of particle number occupancy--operators labeled by position--to 
 justify multiparticle excitations over the vacuum state of the theory on a fixed background spacetime.
Quantization of curved spacetime as a Weyl algebra\cite{Ar-Go_14}
 is consistent with the first notion; indeed, the symplectic structure imposed by the conjugacy of the operators in first quantization leads naturally
 to a {\it derivation rule} among elements of the algebra.
 Incidentally, the sidestepping of coordinate functions in first quantization also provides a fitting path to background independence.

 	In flat spacetime the use of gauge potentials enables a generalization of Newton's notion of infinitesimal differential, the covariant
 derivative, to induce force (vector) fields as a manifestation of geometrical curvature in a sort of ``internal" space with {\it group structure}.
 If this construction sounds like Einstein's interpretation of gravitational force as spacetime curvature, it is because it is inspire by it!
 Yet, rather ironically this edifice turns out to be too simple to address the {\it local} symmetries of general relativity.
 In modern mathematical parlance, such a formalism goes by the name of fiber bundles where the fibers, rooted in the spacetime, carry the internal group structure.
 This construction has its roots in the work of Elie Cartan who also considered the intrinsic problem of spacetime symmetries through a Cartan connection. 
 From a purely formal standpoint \S \ref{sec:CM&NonCenExt}, the spacetime-flat fiber bundle embodied by the ``Poincar\'e space" as a semidirect product
 represents the most perfect fiber bundle!  Indeed, as a smooth fiber bundle it is naturally endowed with the group structure of an affine space.
 Conversely, spacetime-flat fiber bundles with internal symmetries lack the group structure of an enlarged ``group".  
 Cartan's pupil, Charles Ehresmann, is broadly acknowledged for generalizing the notion and popularizing the idea among the physics community 
 by detaching the internal symmetries of the fiber--which correspond to the symmetries of the standard model of particle physics--from 
 the symmetries of the spacetime.    

 	While spacetime proper plays a passive role in Ehresmann's construct--indeed often trivialized as Euclidean 
 space--standard lore furnishes a clever means to accommodate for globally flat spacetime symmetries in the sense of ``spontaneous symmetry breaking".   
 This involves the introduction of a ``principal frame-bundle" (\S \ref{sec:L&R_Acts}): 
 a bundle space where the fibers posses the matrix group structure of {\it general linear maps of Euclidean space}, GL(4,R), 
 followed by a so-called bundle reduction to a sub-group of such linear maps thereby endowing the bundle space with a ``reductive structure".
 With $\cL$ = SO(1,3) the Lorentz group, causal structure is then instilled upon the Euclidean space by 	
 endowing the latter with a reductive $\cL$-structure.  Yet, 		
 the non-existence of a double cover for
GL(4,R)\cite{Kobayashi72, HeylAl95} represents a clear obstruction to the full institution of spacetime structure in frame bundles.

 	Further scrutiny of this picture reveals clear pitfalls when attempting to address the entire anatomy of ``external" symmetries in general relativity.
 Given the universal recognition for the need to formulate quantum gravity in ``coordinate-free" fashion, 
 close scrutiny of the notion of coordinates as {\it sections} of a frame bundle: i.e., functions into local trivializations of the spacetime,
 is clearly warranted.  Our analysis of {\it curved} spacetime highlights an essential distinction between the spacetime proper 
 and its representation in terms of local coordinates on a manifold.
 Perhaps more importantly, a markedly distinct picture of the ``tangent space" to curved spacetime emerges naturally. 
 To clarify the nature of the space and its relation to the Lorentz group, 
 two algebraic differential constructions are examined in \S \ref{sec:CM&NonCenExt}: the standard Fox derivations in a group ring 
 and the more recent left $\alpha$-derivations that accommodate for non-commutative, double-sided vector spaces\cite{Patrick00, Ar-Go_14}.


 	Although the notion of {\it frame bundle} is well adapted to the study of Euclidean geometry, we assert that such an construct as a fibered product 
 is not well adapted to the study of curved spacetime as an abelian surrogate to the Lorentz group \S \ref{sec:CM&NonCenExt}.
 Justification for such a statement lies at the core of our enterprise in this note.
 Furthermore, 	
 we find not only that inclusion of torsion in the framework is necessary for a self-consistent theory but also that incorporation of such an essential piece
 of the puzzle represents an obstruction to the generic trivialization of the tangent space as a fibered product \S \ref{sec:CM&NonCenExt}.
 Remarkably, early models of string theory in the limit of infinite string tension for closed strings; i.e., zero Regge slope, yield non-Riemann, pure torsion
 spacetimes\cite{Sche&Schw74}!  
 Along this line of reasoning, we argue that the tangent space to curved spacetime may only be constructed as a 
 fibered product {\it modulo torsion modules} in the vein of formal homological algebra.  
 Notably, this is consistent with the formal proposal by Artin and Van den Bergh\cite{Art&VDB90} for ``twisted homogeneous coordinate rings" 
 (a natural non-commutative, algebro-geometric generalization of (projective) coordinate functions!) that bear fruit to the notion of 
sheaf bimodules\cite{VDBerg96} and double-sided vector spaces\cite{Patrick00}. 

	 In the following sections we flesh out this story reviewing existing formal constructions of ``G-spaces" 
 while aiming to incorporate the full geometry of curved spacetime symmetries.
 In particular, we introduce the notion that such symmetries may be faithfully represented by
 non-commutative rings in lieu of the Lie groups upon which unification schemes of the standard model are based.
 We will see that even within the standard formal framework of represeantation theory, group theory needs to be supplemented by group rings in order to make sense
 of G-spaces!  Furthermore, if G action on the space is non-trivial, the G-space is decorated by 2-cocyles; a fact that is normally overseen in physical 
 interpretations of spacetime.  The extension from groups to rings within the algebraic cosntruction then seems only natural.
 This paper is a prequel to the technical, mathematical paper that further develops these 
notions\cite{Ar-Go_14}.  Our primary objective here is to build a phenomenological bridge between some hard mathematical facts embedded deep within 
 the technical literature and an old physical problem pursued by the likes of \'Elie Cartan and Albert Einstein.
 More concretely, this note aims to explain how relatively new mathematics 
 may help elucidate ``the gauge status of vierbeins" which along with the spin connection may be interpreted as a byproduct of 
 spontaneous external symmetry breaking of a larger, unified Cartan connection (see \S \ref{sec:AlgEinstein}). 
 Notably, even among the mathematical community the forest is often so dense locally that structures in, say, differential geometry may seem peripherally
 disjoint from their cousins in ring theory.  
 In this short note, a link among these two disciplines will be revealed while embedding spacetime symmetries in the fiber bundle jargon.

 	In \S \ref{sec:L&R_Acts}, 	
 a review is made of the standard fiber bundle formulation while emphasizing on the duties assigned to the left and right actions on the full bundle space
 and their relation to the so-called the gluing conditions for a principal fiber bundle.
 In \S \ref{sec:CM&NonCenExt}, 		
 we take a critical view at abelian group extensions within the framework of formal representation theory and in particular in reference to 
 the Eilenberg-MacLane construction of the classifying space as an abstract simplicial complex.
 Emphasis is placed on the critical notion of Fox derivations in a group ring and it is suggested that these may be replaced by more general
 derivations such as left-$\alpha$ derivations in order to accommodate for spacetime bundle structure.
 At last, in \S \ref{sec:AlgEinstein}  
 an algebraic analysis of Einstein's equation is performed in order to gain insight into a plausible construction of the 
 {\it dynamical curved spacetime} in the vein of the formal machinery laid out in \S \ref{sec:CM&NonCenExt}.

\section{Left and Right actions on Principal Fiber Bundles} 
 \label{sec:L&R_Acts}

	Locally, the full bundle space of a fiber bundle may be taken as a ``Cartesian product" of base and fiber: E$_{\rm local} \equiv$ {M $\times$ F}.  
 A ``local trivialization" of such a space is,
loosely speaking\footnote{ 
\label{FN:LocTriv}
Local trivializations are actually maps from ``points" in the bundle space to local coordinates.
Given a partition of unity on the base manifold as a topological space and charts subordinate to the open cover indexed by such a partition,
\[
	\psi_i:U_i \times H \rightarrow \pi^{-1} (U_i),
\]
each such diffeomorphism as {\it inclusion} into the fiber bundle must satisfy {\bf right equivariance} on the fibers:
$\psi_i (u_i, h_i) = \psi_i (u_i, e) h_i$ so that {\it a given chart corresponds to a local section} $\sigma$ over $U_i$: $\psi (u_i, h) = \sigma(u)h$.
On overlaps, two sections from different trivializations must agree on the fiber
\[
	\psi_i \rightarrow \pi^{-1} (U_i) = \sigma_i(u_{ij}) \, h_i = \sigma_j (u_{ji}) \, h_j  = \pi^{-1} (U_j) \leftarrow \psi_j
\]
so that $\sigma_i$ and $\sigma_j$ are related by a smooth map, k: $U_{ij} \rightarrow \cH$, acting on the {\it right} of sections: 
$\sigma_j = \sigma_i k$ or 
\[
k = \sigma_i^{-1} \sigma_j |_{u_{ij}}.
\]
Thus, composition of sections on overlaps is in one to one correspondence with the transition functions and one can {\it reconstruct} the bundle 
from transition maps. 		
The crucial key in this result is the existence of an ``equivariant structure" on the underlying bundle space
since such structure enables the use of a ``zero section" on a local patch, $U_i$, to span the vertical space {\it via} right translation along the fibers:
\[
 \psi_i (u_i, h_i) = \psi_i (u_i, e) h_i  =  \sigma_0(u_i) h_i.
\]
}, given by local coordinates $(a,f)$ 
 where $a \in $ M is a coordinate on the ``spacetime" and $f \in $ F, a coordinate on the fiber; i.e., on an internal space that does not necessarily  
 possess group structure. 
 The precise definition of local trivialization is cumbersome but necessary to formalize the framework.  
 However, 
 for the sake of clarity in this section the formal development will be carried out in parallel fashion through the footnotes.

 	Additional data attached to the fiber bundle is a {\it Lie group}, G, which acts on the left of {\it the full bundle space} 
 inducing a change of trivialization (i.e., change of local coordinates): 
\[
	g \cdot (a, f)  = (a,gf) \equiv (a, f'). 
\] Notably, the base (spacetime) coordinates are not affected by the (left) group action; one then says that the group action preserves the fibers pointwise.
 To the physicist, the notion
 of symmetry group and group manifold are intertwined in a somewhat mysterious way.  
 Within the context of Lie groups, ``points" on a group manifold--which is {\it locally} modeled as Euclidean space--are 
 tacitly and commonly identified with group elements as (not necess. differential) operators on the manifold and 
vice versa\cite{Lord&Gosw88}.
 On the other hand, within the realm of {\it formal representation theory} a Lie group is {\it both}, a symmetry group, call it G, and a ``G-vector space":
 a {\it left ZG-module} resulting from the generation of a unital, associative, graded {\it ``integral group ring"} in a canonically defined way.   
 The reader will no doubt notice the abrupt introduction of a ring structure in the framework; 
 this is essentially a fancy way to state that in order to work with the group one needs to construct a ``group space" and this is accomplished 
 in the spirit of a {\bf topological Taylor series expansion}.  Pictorially, this may be portrayed {\it via} the so-called ``bar notation\cite{Eil&McL53}":
\[
	B(G) = {\bOne}_{_{\bf Z}} + G + G|G + G|G|G + ...
\]
 More technically, 
 the components of the ``classifying space" {\bf BG} are sums of {\it simplex chains} with vertices made up of (additive) group elements up to some degree
 and with coefficients in the ring of integers Z.   
 This level of abstraction is not essential to this note but this picture is worth a great many beautiful mathematical thoughts. 
 Let us begin to frame the semantics involved in the interpretation of such a depiction.

 	In a minimalist way, a group is a set with (two-sided) identity and a single invertible operation: addition or multiplication.
 On the other hand, a ring looks like a mixed algebraic structure laying between additive and multiplicative groups,
 naturally endowed with both binary operations and with multiplication distributing over addition.
 Despite first impressions, rings are in fact {\it algebraically more primitive} than either groups or fields. 
 A working definition of a group in mathematical physics necessitates an augmentation of its abstract definition with a ``space" for the group to ``act upon".
 Furthermore, such a space is typically built ``over" a commutative ring or a field, i.e., a division ring with multiplicative inverse.
 Now the hybrid notion of a group ring is interesting as a formal proposal on how to build a {\it polynomial ring} from both a ring and a group:
 by allowing coefficients in the polynomial to take values on a ``ground ring"--the ring of integers Z, above--while
 ``indeterminates" take values on the elements of the group G.  
 Cornerstone to this construction are the (multiplicative) unit elements in both the ring and the group
 which together--the ring unit on the left multiplying the group unit on the right--constitute the unit as the starting point in the Taylor 
 expansion as a polynomial.		
 Note, nevertheless, that just like in the Taylor expansion of the 
 sin$(\theta - \theta_0)$ there is no preferred choice of origin, $\theta_0,$ to expand about. 
 We will see below that {\it derivations} in such a polynomial with unit bridge the notions of group and ring in a {\it universal} albeit asymmetric manner. 

	Back to physics, 
 making the distinction between group elements: $g \in$ G, and points in the manifold, $a \in$ M, 
 one can define left and right actions $L_g \equiv ga$ and $R_g \equiv ag$, as left and right {\it translations} on M.  
 The canonical choice of action for a Lie group is left action which gives rise to the crucial notion of left invariant vector fields 
 as elements of the Lie algebra {\it via} the {\it Maurer-Cartan 1-form} (see below).
 It is rather interesting that working definitions of such notions could in principle be built from the right action as well;
 yet, left and right constructions are {\it not} equivalent when the topology of the group is non-compact (e.g., for the Lorentz group).
 Still, if the fiber and the group are the ``same" as group manifolds (i.e., isomorphic: F $\simeq$ G),
 then the group may act on the right of the ``full bundle space" as well:
\[
	(x, g) \cdot h = (x,gh), 
\]
 and this action makes it possible to {\it parallel transport} vector fields and to lift horizontal curves to the bundle space in a self-consistent manner
 even when the structure group is non-compact (see, e.g., Proposition 6.36 of Morita\cite{Morita97}).		
 In fact, in this case one speaks of a smooth, ``principal" fiber bundle where the right action is independent of the choice of 
local trivialization\footnote{On overlaps:
\[
	uh = \psi_j(x ,g_jh) = \psi_j(x, t_{ji} g_i h) = \psi_i(x ,g_i h),
\] which demonstrates that right action is independent of the local choice of local trivialization.	
}
 and horizontal paths are simply lifted by such an action.  
 More technically, a principal fiber bundle is endowed with an ``equivariant"
structure\footnote{
\label{FN:EquivStruct}
Define, $\iota_x$ as the fiber part of a local trivialization, a {\it ``diffeomorphism"} 
over an open neighborhood of $x \in M:~ \iota :~ U \times \cH \rightarrow \pi^{-1} (U)$,
\[ \iota :~ U \times \cH \rightarrow \pi^{-1} (U), ~~~~\iota^{-1} (p) = ( x, \iota^{-1}_x(\pi^{-1} x) ) = (\pi, \iota^{-1}) \, (p)|_x \longmapsto (x,h), \]
consistent with right translation spanning the fiber space: $\iota_x(hg) = \iota_x(h) g$ 
and {\it invariant} by the left composite diffeomorphism induced by a change of trivialization on overlaps: 
$(\iota_x \circ L_{g})^* \Theta = \iota_x^* \Theta \equiv \omega$, 
as applied to the Ehresmann connection, $\Theta$, whose ``pullback" in turn yields the (left invariant) Maurer-Cartan form, $\omega$, 
valued on the Lie algebra of the fiber.
}: a property of local trivializations consistent with right translation spanning the fiber space
 and such that the differential structure of the fiber space remains {\it invariant} by 
 the left translation induced by a change of trivialization.  
 At a deeper level, the equivariant structure in fact also imposes a condition on the base space: it must be invariant by the right action of the 
 structure group\cite{Ar-Go_14}.
 This enables the ``gluing" conditions that permit the fiber bundle to be ``put together" pointwise fiber by fiber throughout the base space
 and provides formal motivation to the statement that the structure group preserves the fibers pointwise.
 In addition, this is also related to the anti-podism that enables the homology and cohomology of groups to be built from left G-modules
 alone (this is explicated thoroughly at the end of \S \ref{sec:CM&NonCenExt}).

	$\cL$eft invariance of such a ``vertical" space is pivotal to the structure of groups and fiber bundles.  
 To understand this notion, one must first make a distinction between the differential space at an arbitrary element of the group 
 a.k.a. the tangent space attached to each ``point" of the group manifold
and the differential space at the identity element. 
 For finite dimensional groups, both spaces are spanned by elements of the {\it Lie algebra}: 
 roughly speaking given by {\it infinitesimal measures} of the difference between right and left translations on a differential group element. 
 A {\it linear functional} on elements of the Lie algebra that live at the space tangential to an element of the group, T$_g$,
 which ``brings them back" to the space tangential to the identity element, T$_e$, defines the Maurer-Cartan one-form $\omega_B$.
 Informally, one may say that the one-form $\omega_B$ (acting on elements of the Lie algebra) is the {\it pull back} to the Id element.
 Furthermore, to extend this construction from the Lie algebra of the fiber to the entire bundle space one makes use of the equivariant structure introduced in 
 the previous paragraph (i.e., footnote[\ref{FN:EquivStruct}]). 

\section{Crossed Modules and Non-central Extensions} 
 \label{sec:CM&NonCenExt}

 	Although seldom fully appreciated; there are several roughly equivalent ways to describe smooth fiber bundles:
 semi-direct products, group extensions, short exact sequences, fibered products, crossed modules, 2-groups etc.  
 In this section, we take a critical view at these mathematical structures, their implicit assumptions and their embeddings into universal constructions 
 aiming to elucidate their use in the context of curved spacetime bundles. 	
 We will only be concerned with bundles where the full bundle space results from a {\it natural embedding of the spacetime} as defined below.
 This in turn means that the full bundle space may be broadly interpreted as a ``group manifold" or as an appropriate categorification of such a notion.
 Furthermore, adhering to the general goal of this note, we will begin with the simplest notion and build our way up to the more sophisticated descriptions.

	By a natural embedding of a group into another we will mean that the former group, call it N, 
 may be viewed as a {\it normal} sub-group of the latter one, call it E.
 In this case, E acts naturally on N by conjugation: $ene^{-1} \in N$ and the group quotient of E by N is naturally endowed with group structure.
 In particular, 
 elements of the quotient are well parsed into {\it equivalent classes} which define ``points" of the quotient as a group (under the quotient topology)
 with a well defined product operation;
 i.e., a binary operation between elements of the equivalent classes that is independent of the choice of [coset] representative for each class.
 A {\it homogeneous model geometry} is the quotient of a ``large" group by a subgroup that is not necessarily naturally embedded.  
 Thus, if the latter is not a normal subgroup the quotient is loosely labeled a ``vector space" but this space is devoid of a binary group operation rule.
 Four dimensional Euclidean space, R$^4$, represents a canonical trivialization of what a physicist refers to as a vector space.
 In special relativity,
 R$^4$ is equipped with some minimal extra structure to enforce causality through an inner product $\langle \bf{a}, \bf{b} \rangle_\eta = \bf{a}^\dagger \eta \bf{b}$
 defined {\it via} a globally flat metric: $\eta = $ diag(-1,1,1,1): \Mink $\,\equiv$ (R$^{1+3},~ \langle, \rangle_\eta$).
 In general relativity, the ``world metric", $g_{\mu\nu} = e_\mu^a \,\eta_{ab}\, e^b_\nu$, is {\it locally} derived from the flat metric 
 through the introduction of {\it vierbeins} (a.k.a. frame fields), 
 $e_\mu^a$, which possess one tangent, globally flat index $a$ and one world, locally curved spacetime index $\mu$ (\S \ref{sec:AlgEinstein}). 
 Thus, vierbeins may be informally thought of as the ``square-root of the metric" whereas the 
 Minkowski metric is attributed to an ``internal", tangent space which is globally flat by definition.

 
	In its simplest but perhaps most profound form, the semidirect product
 of two groups requires the first group, call it A for abelian, to be surrogate to the second, call it G for group. 
 Thus, A must be a G-vector space or a {\it symmetric ZG-bimodule} to be precise:
 a vector space with coefficients in the integral group ring {\bf ZG} where right and left multiplications by elements of {\bf ZG} are 
 equivalent {\it via} the existence of an 
{\it ``antipodism"}\footnote{
\label{FN:AntiPode}
Technically, the existence of an antipode in the group multiplication rule allows us to axiomatize this operation as commutative diagrams 
in the fashion of a Hoft (bi)algebra.  Furthermore, 
given {\bf ZG} as a ring, a left ZG-module is naturally equipped with an ``anti-automorphism" that allows for its interpretation as a right module.
In particular, 
left action by $g \in ZG$ on A as a left ZG-module: $ga$, is equivalent to a right action by $g^{-1}$ on A as a right ZG-module: $ag^{-1} \simeq ga$
so that in the case of groups, one can generically make sense of the tensor product of two left ZG-modules.
This is the meaning we attach to the term  ``symmetric bimodule" above. 
Note that this interpretation is formally consistent with the scheme-theoretic notions introduced by Artin and Van den Bergh\cite{Art&VDB90}
for locally free $\cO_X$-bimodules over {\it generic points}.
}. Again, 
 this level of abstraction is not essential to our aim here but the ``antipode" notion will resurface again below.
 The semidirect product is then broadly defined as an {\it asymmetric Cartesian product}, E $\equiv$ A$\sdprod$G, 
 naturally endowed with an asymmetric group multiplication law:
\[ (a,g)\cdot(b,h) = (a+gb, gh), ~~a,b \in {\rm A}, ~ g,h \in {\rm G}
\] consistent with the ``left" action of G on A as a G-space.
 If such an action is trivial, G$\cdot a = a$, the semidirect product yields the simplest form of 			
 group extension: a {\it central group extension} of $G$ by $A$ where, by definition, all the elements of A commute with all the elements of the extension.  
 Notably, it is the nature of the action on the G-space that bears the type of group extension; e.g. central {\it vs} non-central extension.

	Intriguingly, the asymmetry in this group concatenation may be gracefully construed as left action by an {\it affine (symmetric) bimodule-group tuple}
 in the spirit of the left group action on a vector bundle.  
 Not surprisingly, the fiber part of such an action is multiplicative and left-right symmetric:  $(\UL ,g)\cdot(\UL ,h) = (\UL,gh)$.
 Far more revealing is the full left action on the bimodule structure to the right for it does not preserve the fibers pointwise.
 Indeed, left action by $(a,g)$ on $(b,\UL)$ induces both a left group action plus an additive translation of the origin: $(a,g)\cdot(b, \UL) = (a+gb, \UL)$.
 Let us indulge on whether this so-called ``asymmetry" in the group concatenation is truly so.
 Concretely, can we interpret the multiplication law as right action on the geometrical object to the left?  
 The answer is yes: by means of the proper mathematical jargon this object is a {\it right-torsor--group tuple}.
 The fiber part of such an action is simply given by right group multiplication just as in a {\it principal} fiber bundle.
 On the other hand,
 the right G-torsor action of $(b, \UL)$ on the torsor-group tuple to the left: $(a,g) \cdot (b, \UL)$,
 requires that $b$ act on $g$--{\it via} the G-action--before it gets to act additively on {\it a} as  a torsor element.
 Of course, the full action yields an identical result: $(a,g)\cdot(b,h) = (a+gb, gh)$. 
 In the literature this is known as a non-linear transformation of a (flat) fiber bundle.  
 Granted that our general goal is to extend the fiber bundle construct to curved space, 
 this link is rather noteworthy; we will duel on it's mathematical origins a bit.

 	Informally, one may justify these concatenations as motions of geometrical objects past each other {\it via} predefined algebraic operations; 
 a parody which is formalized below.  
 The added perk is that of blurring the notions left and right actions into a single associative group multiplication rule.
 The key conceptual difference of a G-torsor as an affine bimodule is that it does not posses a preferred choice of origin
 just like the function sin$(\theta)$ may be realized for any value of $\theta$ without examining its local differential structure at a preferred 
 value of $\theta$.
 
	Proper justification of the group multiplication law in an abelian group extension requires a close examination of the embeddings involved.
 Recall that, as a normal embedding, E acts on A by conjugation: $\tilde{g} i(a) \tilde{g}^{-1}$ where, following Brown\cite{Brown82}, 
 $\tilde{g} \in E$ is the image of $g \in G$ in $E$ 
 and $i(a) \in E$ is the injection of $a$ into $E$.  
 Now, given that A is abelian, its elements commute with themselves as a subgroup of the E; e.g.,  $i(a)i(b) = i(b)i(a)$, 
 so the group quotient of E by A induces an action of G on A whose image resides in A and whose image in E is $i(ga)$, 
 where $g = \pi(\tilde{g})$ with $\pi: E \rightarrow G$ defining a projection map.  
 Assembling these two statements results in
\[ i(ga) =  \tilde{g} i(a) \tilde{g}^{-1} ~~{\rm or,~written~as~a~commutation~rule\cite{Brown82}~in~E:}~~ i(ga) \tilde{g} =  \tilde{g} i(a)
\]
 so an {\it abelian} extension implies an action of G on A.  As usual, the canonical choice of action is left action.
 These observations have deep implications for the representation theory of groups and smooth fiber bundles in general. 
 Let us elaborate on the algebraic manipulations involved as these shed light into the formal construction.

	A split group extension--a.k.a. a {\it crossed homomorphism}--is an abelian extension for which the exists a ``splitting" map:
 $\sigma: G \rightarrow E$ such that $\pi \circ \sigma \equiv \bOne_{_G}$.
 Critical to our discussion here is the fact that a splitting of the extension is equivalent to the existence of 
 a {\it canonical choice} of local section $\sigma: G \rightarrow E$, in the lexicon of fiber bundles (Ref. Footnote [\ref{FN:LocTriv}]). 
 Superficially, this basically means that if one introduces coordinates on A as a ``fiber space" there is an unambiguous choice of origin 
 on each local trivialization.  More profoundly, the failure of a section to conform a {\it group homomorphism} is intrinsically related to the 
 differential structure of the G-space as a proxy for G.

Aside from technical details\footnote{
\label{FN:Non-Split_GE} 
In fact, when the action of G on A is non-trivial the section $\sigma$ ``fails" as a homomorphism: $\sigma(g)\sigma(h) = i(f(g,h) \sigma(gh)$ 
which is parametrized by a {\it ``factor set"}, $f$, defined as an element of A, $f\!\!: G \times G \rightarrow A$.
This, in turn, adds a term to the affine part of the group multiplication law:
$(a,g) \cdot (b,h) = (a + gb + f,~ gh)$, 
and changing the choice of section corresponds to modifying the factor set by a coboundary\cite{Brown82}:
$\delta c = c(g) + g c(h) - c(gh)$; i.e., by the image of the derivation map which in this case is a Fox derivative.
}, 
 splittings are {\it derivations} in a group ring:	 
\[ \sigma: G \rightarrow E ~~|~~ \sigma(g) \equiv \tilde{g} = (Dg, g).
\] 
 Generally, derivations are ``objects", $D$, obeying the Leibniz rule:
 $D ab = D(a) b + a D(b)$; however, derivations in a group ring obey the so-called 
Fox\cite{Fox53} free derivative rule instead:
 $D gh = D(g) 	+ g D(h)$.
 This somewhat mysterious behavior has roots in the topological bar construction outlined above.
 Algebraically, however, we may justify it as follows. 
 Given the maps $i$ and $\sigma$, write a general element of $E$ as $i(a)\sigma(g)$.  
 We may now write a product of two elements of E: $i(a)\sigma(g) i(b) \sigma(h)$
 and use the commutation rule to commute the middle terms thus bearing the celebrated product rule for an abelian group extension:
 $i(a) i(gb) \sigma(g) \sigma(h) = i(a+ gb) \sigma(gh) \equiv (a+ gb, gh)$.
 Now, write $\sigma(g) = (Dg, g)$ and $\sigma(h) = (Dh, h)$ and compute the product $\sigma(gh) = (Dg, g) \cdot (Dh, h) = (Dg + g Dh, gh)$. 
 Noting that $\sigma(gh) = (D(gh), gh)$, the enigmatic Fox derivative rule follows.

	Let us now take a close look at the root of the commutation rule that unraveled such far reaching ramifications.  
 Using the ``set theoretic bijection"\cite{Brown82} between elements in the pair (A, G) and elements of $E$: $(a,g) \longleftrightarrow i(a)\sigma(g)$, 
 the adjoint action of E on A is given by: $i(a)\sigma(g) i(b) \sigma^{_{-1}}(g) i^{_{-1}}(a)$.
 The middle elements need to be commuted under the premise that G acts on A by induced action. 
 There are {\it two} choices leading to either left or right action on A.  
 Commuting $\sigma(g) i(b)$ yields $i(gb)$ as an end result whereas commuting $i(b) \sigma^{_{-1}}(g)$ yields $i(bg^{-1})$.
 This equivalence between right and left actions: $[ga] \sim [ag^{-1}]$, motivates the ``antipodism" referred to in Footnote [\ref{FN:AntiPode}]. 
 It is rather noteworthy that such an {\it equivalence class} between right and left actions enables 
 the homology and cohomology theory of groups to be developed from left G-modules exclusively\cite{Car&Eil56}.

	Taken at face value, the Fox derivative rule implies a strong constraint on the G-space.
 Indeed, regardless of the nature of the left action the space must be invariant by the right G-action 
 (of course if one chooses the right action as the canonical action, the space must be invariant by the left G-action instead).
 This makes the canonical G-space an asymmetric bimodule. 
 We will christen the G-space as a {\it double-sided vector space} to conform with the recent formulation of
David Patrick\cite{Patrick00}.  
 Next, the root of this asymmetry is briefly unraveled.

 	Recall that the Fox derivative occurs within the context of formal representation theory.  
 An element of the integral group ring, ZG, is a finite sum of group elements: $\sum a_i g_i$, with coefficients $a_i$ in the ring of integers Z.
 ZG is naturally endowed with an ``augmentation map" into the ring of integers, $\epsilon: ZG \rightarrow Z$, as a {\it surjective ring epimorphism}.	 
 This is simply achieved by setting all $g_i \rightarrow 1$ for any element in the ``free" polynomial ring generated by the pairs ($a_i, g_i$). 
 The kernel of this map--call it IG for the ideal of ZG corresponding to G itself--amounts to elements of coefficient sum zero\cite{Fox53}. 
 Fox derivatives, $\tilde{d}: ZG \rightarrow A$ are (extended) derivations of elements of ZG:
 $\tilde{d} (rs) = r \tilde{d} (s) + \tilde{d} (r) \cdot \epsilon(s)$,
 which, when restricted to act on G (as a subgroup of ZG) yield the Fox derivative rule on the group alone.   
 It is easy to see that for derivations in the full integral group ring 
 the {\it right} group action is replaced by the image of the augmentation map on elements of ZG.
 Generally, a given derivation rule will lead to a {\it differential polynomial ring} in lieu of a crossed homomorphism.

\section{An Algebraic Analysis of Einstein's Equation}
\label{sec:AlgEinstein}

{\footnotesize
\begin{flushright} 
{\it
``We could still imagine that there is a set of laws that determines events completely for some supernatural being,
who could observe the present state of the universe without disturbing it.  
However, such models of the universe are not of much interest to us mortals.  
It seems better to employ the principle known as {\it Occam's~razor} and cut out all the features of the theory that cannot be observed." } \\  
Stephen Hawking (circa 1988).
\end{flushright}
}

 	The moral rooted in \S \ref{sec:CM&NonCenExt} is that the G-space must be {\it derived} from the group G in a canonical way {\it via} a suitable derivation
 rule.  Furthermore, when the group acts on the space non-trivially it decorates the space with geometrical structure contingent upon the adopted derivation rule
 (Ref. footnote [\ref{FN:Non-Split_GE}]).
 Thus, abelian extensions which do not conform to a derivation rule should not be considered ``natural" embeddings in any physical way. 
 This is a rather robust statement which nevertheless is in juxtaposition with standard lore where the Lorentz group is informally extended by Euclidean space
 or by its automorphism group: GL(4,R). 
 On the other hand, the well known problems\cite{Kobayashi72, HeylAl95} 
 in trying to embed the Lorentz group $\cL$ (and in particular its double cover $\bar{\cL}$) in GL(4,R) are gracefully circumvented by the bar construction. 
 In {\it flat} spacetime, 
na\"ive\footnote{ 
 Reader beware of the fundamental underlying assumption in the structure of the classifying space: BG is {\it contractible}, 
 i.e., with fundamental group isomorphic to G and trivial higher homotopy groups.  
 Evidently this construction needs elaboration to accommodate for the Lorentz group's considerable complexity: non-compactness, discrete symmetries, etc.
 Yet, we take this as a first step towards our understanding of curved spacetime as a {\it dynamical ring}.
} application of the bar construction to the Lorentz group yields the Poincar\'e group as an affine space
 decorated with 2-cocycles: as explained above, the G-space constitutes a {\it right G-torsor} to be precise. 
 Let us emphasize that the resulting space is flat; this is known in mathematics as a {\it holomorph}:
 the Lorentz group acts on the $\cL$-space by automorphisms. 

 	To shed physical insight on the nature of the {\it curved} $\cL$-space, let's indulge for a moment in the following {\it gedankenexperiment}.
 Inside Einstein's elevator, Lukas measures random inertial forces with an accelerometer.  
 These may be caused either by random motion of the elevator with respect to an inertial frame attached to an external body of infinite mass 
 (so that the frame is ``fixed" with respect to distant stars)
 {\it or} 
 by random external bodies--massive ``blobs"--flying past the elevator with mass distributions to match the measurements.
 By the equivalence principle, Lukas is unable to distinguish between these two scenarios.
 Yet, our observer is a well equipped engineer endowed with a watch and a meter stick thereby declaring his spacetime to be metric and causal. 
 Utilizing some formal semantics now general covariance of physical laws in Lukas' frame is manifested by the fact that his readings may be reproduced by random 
 motions of the elevator with respect to a ``universal frame" as erected above.  On the other hand, 
 the equivalent readings obtained by {\it causally} flying blobs {\it necessarily} endow the ``external spacetime" with causal structure as well. 
 Thus, the spacetime is causal throughout and 
 {\it diffeomorphism invariance is restricted to frame transformations derived from local Lorentz transformations}
 very much in accordance with the bar construction as advertised in \S \ref{sec:CM&NonCenExt}.
 Quite remarkably, Einstein's equation also suggests a way to generate the {\it curved} spacetime algebraically as it relates the field strength of rotations 
 with Minkowski signature--the Riemann curvature--to the current of translations--the {\it non-canonical} stress-energy tensor.
 Note that in this interpretation, the curvature is valued in the Lie algebra of $\bar{\cL}$ which we denote by $\bar{\cL'}$ 
 while the current takes values on its {\it ``derivation space":} $D\bar{\cL}$.

{\footnotesize
\begin{flushright} {\it --Pluralitas non est ponenda sine neccesitate--} \end{flushright}
} 

 	{\it$\cO\!$ur} starting point is a brief analysis of a slightly relaxed version of the Palatini action (without cosmological constant):
\beq
	S^P (\omega, e) = \int_M  e^\mu_a e^\nu_b \;  \cR^{ab}_{\mu\nu}  \, {\rm Vol}  = \int_M  \epsilon_{abcd} \; e^a \wedge e^b \wedge \cR^{cd}
\label{Eq:ModPalat}
\eeq		
 where the ``extended Ricci scalar", $e \,\cR\, e \equiv e \, (\cS + \cT) \, e$, is {\it naturally} endowed with a Lie~algebra-derivation~space decomposition;
 i.e., as pure curvature, $\cS \equiv \cD \omega$, and pure torsion, $\cT \equiv \cD e$, components respectively 
 (with $\cD = d + \omega$ the ``spin" covariant derivative) 
 and
 where both the connection 1-form (a.k.a. spin connection), $\omega$, and the vierbein 1-form, $e$, are considered as {\it independent} variational fields.
 The middle expression is easy to work with and we make extensive use of it below.
 It conforms with the standard mathematical structure of the action as a Lagrangian {\it density} integrated over a spacetime {\it volume} Vol.
 In general relativity, the latter is commonly expressed as Vol = $\sqrt{-|g|} \; d^4x$ whereas in terms of the vierbein 
 one has Vol = $\half |e| d^4x$ where $|\UL|$ stands for the determinant. 
 Conversely, the expression on r.h.s. is not only a more elegant way to express the gravitational action but also explicitly manifests
 the {\it intrinsic} geometrical meaning of the gravitational Lagrangian as a 4-form:
 neither a pure density nor a pure volume form but rather as the intertwiner of the {\it total} curvature 2-form and two vierbein 1-forms.
 
	To make contact with Einstein's general relativity, we simply observe that the Lie algebra-valued component of the curvature 2-form,
 $\cS^{ab}$ (tangent space indices), and the Riemann curvature tensor, ${\rm R}^\alpha_{\;\beta\mu\nu}$ (world indices), 
 are related through:
\[ 	\cS^{ab} ~~(= \cS^{ab}_{~~~ \mu\nu} dx^\mu dx^\nu ) 
 \equiv 
	\half {\rm R}^{\alpha\beta} e^a_\alpha e^b_\beta 
	~~(= \half {\rm R}^{\alpha\beta}_{~~~ \mu\nu} e^a_\alpha e^b_\beta \; dx^\mu dx^\nu )
\] where the ``world differentials" $dx^\mu$ are often hidden for convenience in notation. 
 If the {\it total} curvature (Ricci) ``scalar",
 $e^\mu_a e^\nu_b \,  \cR^{ab}_{\mu\nu}$, is restricted to take values on the Lie algebra of $\bar{\cL}$ excluding its derivation space, $D \bar{\cL}$
 (and thus the torsion component), the standard form of the Hilbert action obtains. 	

	Cornerstone in this formulation is the existence of two independent {\it dynamical} variables, $\omega$ and $e$, instead of the metric alone: 
 g$_{\mu\nu} = \eta_{ab} e^a_\mu e^b_\nu$, with $\eta$ the non-dynamical, globally flat Minkowski metric on the tangent space 
 (for the time being, we write ``$e$" for both $e$ and $e^{-1}$ assuming of course that these are freely invertible throughout the space).
 At a very fundamental level, this is consistent with the interpretation of the {\it sum} of spin connection and vierbein as a {\it Cartan connection}:
 $\zeta = \omega + e$, taking values in the algebra of the entire bundle space: 
 $\cZ' \equiv \bar{\cL'}  + D \bar{\cL}$, 		
 where primes denote algebras and where we invoked the double cover of the Lorentz group, $\bar{\cL}$,
 to instill spin structure upon the curved $\bar{\cL}$-space (of course, with the $\bar{\cL}$-space as a {\it derivation space} $D \bar{\cL}$).
 Notably, the curvature of a Cartan connection, $\cD\zeta$, takes values on the algebra of the entire fiber bundle
 (which corresponds to the Poincar\'e algebra in the limit of zero Cartan curvature).  
 In general, this curvature does include a torsion component as the 
 $D \bar{\cL'}$-valued part of $\cD \zeta$ (pg. 184 of Sharpe\cite{Sharpe97})
 but this vanishes when the curvature takes values only on the Lie algebra of the generating (non-abelian) group $\bar{\cL}$
 (reader beware of the difference between the covariant derivative $\cD = d + \omega$ and the algebraic derivation $D$; e.g., the Fox derivation).

 	It is well known that in the limit of zero torsion ($\cD \zeta \in \bar{\cL'}$), setting
 $\delta S^P|_{\omega= {\rm const.}} = 0$ and $\delta S^P|_{e= {\rm const.}} = 0$ independently yields the expression on the l.h.s. of Einstein's equation: 
 the symmetric, world-indexed Einstein's tensor $G_{\mu\nu} = {\cR}_{{\mu\nu}} - {1 \over 2} {\cR} {\rm g}_{{\mu\nu}}$, from variation of the Palatini action 
 alone.  Basically an {\it asymmetric} version of $G_{\mu\nu}$ is obtained from setting $\delta S^P|_\omega = 0$ while
 $\delta S^P|_e = 0$ symmetrizes the Levi-Civita connection and thus $G_{\mu\nu}$ as well (see, e.g., {\it ``Gauge Fields, Knots and Gravity"}\cite{Baez&Muni94}).
 On the other hand, motivated by a wish to interpret the r.h.s. of Einstein's equation as a {\it conserved current} in the spirit of Emmy Noether's theorem,
 relaxing the no-torsion condition will lead to asymmetric tensors instead.
 Note that although the Lagrangian density mixes $e$, $\omega$ and their covariant derivatives, $\cT = \cD e$ and $\cS = \cD \omega$, 
 independent variation by each of these, $\delta S|_e$ and $\delta S|_\omega$ defines a natural decomposition into two separate equations with values on 
 the Lie algebra of the group and on the derivation space respectively.
 In the matter sector, it would then be quite {\it natural} to define the stress-energy and spin currents:
 T $= \delta S^M| _{e= {\rm const}}$ and $\cJ = \delta S^M| _{\omega= {\rm const}}$, respectively.


	Although abiding by Occam's razor principle this seems like a reasonable plan to follow, inclusion of torsion in the action proves to be rather elusive.
 Variation of the extended action with respect to the vierbein while holding $\omega$ constant yields:
\bea
\label{Eq:AG_Act}
	\delta S|_{\omega = {\rm const}} &=& \int_M \delta (e^\mu_a e^\nu_b )  \cR^{ab}_{\mu\nu}  \, {\rm Vol} 
					 +  (e^\mu_a e^\nu_b )  \delta \cR^{ab}_{\mu\nu} \, {\rm Vol} 
					 +  (e^\mu_a e^\nu_b )  \cR^{ab}_{\mu\nu} \delta {\rm Vol}			 		\cr 
					&&\cr
					 &=& \int_M \left[ 2(\delta e^\mu_a e^\nu_b)  \cR^{ab}_{\mu\nu}  
					 +  (e^\mu_a e^\nu_b )  \delta \cT^{a}_{\mu\nu} 
					 -  (e^\kappa_c e^\nu_b )  \cR^{cb}_{\kappa\nu} e^a_\mu (\delta e^\mu_a) \right] \, {\rm Vol} 	\cr 
					&&\cr
					 &=& 2 \int_M \left\{ \left[ e^\nu_b \cR^{ab}_{\mu\nu}  
					 -  \half (e^\kappa_c e^\nu_b )  \cR^{cb}_{\kappa\nu} e^a_\mu \right] \delta e^\mu_a
					 +  \half (e^\mu_a e^\nu_b )  \delta \cT^{a}_{\mu\nu} \right\} \, {\rm Vol}, 	
\eea
 where the symmetry properties of $\cR: ~ \cR^{ab}_{\mu\nu} = -\cR^{ba}_{\mu\nu} = -\cR^{ab}_{\nu\mu}$ 
 and the identity $\delta$Vol = $-e^a_\mu (\delta e^\mu_a)$Vol are used in the second line to tidily factorize terms.
 The alert reader will notice a subtle mismatch in geometric objects as evidenced by the index structure of the terms involving $\cT$ in Eq [\ref{Eq:AG_Act}].  
 In a nutshell, na\"ive inclusion of a torsion term in the extended Palatini action is problematic in that it spoils the compactness of the Hilbert action 
 as an integral of a density over a volume element.  By itself this should not be surprising: 
 as emphasized above the description of the gravitational action with such structure is artificial!   
 Notwithstanding the expected shortcomings, our attempt at incorporating torsion in the action at a fundamental level reveals a deeper truth: 
 Introducing torsion represents an obstruction to the ``nice" gluing conditions that enable the construction of general fibered products or crossed-homomorphisms  
 as smooth, principal fiber bundles in flat spacetimes within the standard group theoretic framework.  
 We comment briefly on the ``homological" implications of the existence of torsion in the concluding remarks while immediately occupying ourselves 
 with a more geometrical explication of torsion in hopes that such analysis will shed some physical intuition on why ``The~Legos~Do~Not~Fit"
 when torsion is included in this game.

	From a purely differential geometric point of view, the fundamental object of interest here is the vierbein, \be, as a frame field 
 (with dual 1-form: $\bvarep$).  Differential structure is instilled upon the spacetime {\it via} two geometric objects: 
 the covariant derivative, $\Del_{\UL}\,\UL$, {\it and} the Lie derivative $\bcL_{\UL}\,\UL = [\UL, \UL]$.
 Both of these have {\it intrinsic} meaning as abstract standalone symbols defined largely by
linearity\footnote{
 $\Del$ is the {\it tangential} component of the derivative of a vector {\it field} \bv--as an argument of the $\Del$ operator--along 
 a given vector $\bX$ defined at any given point of the space $p$: $\Del_\bX \bv$.  
 Since there is no preferred point in the space, this derivation constitutes an affine connection.
 Acting on a smooth-pointwise-function--vector-fields amalgam, $f\bv$, 
 both the affine connection and the Lie derivative are defined fundamentally by {\it linearity} rules:
 distributivity under addition and linearity under scalar multiplication.  For instance, $\Del$ obeys
{\it i-} $\Del_\bX \, (a\bv + b\bw) = a\Del_\bX \bv + b\Del_\bX \bw$,~
{\it ii-} $\Del_{a\bX + b\bY} \, \bv = a\Del_\bX \bv + b\Del_\bY \bv$, plus linearity under a ``composite" form of Leibniz rule:
{\it iii-} $\Del_\bX (f\bv) = \bcL_\bX(f) \bv + f \Del_\bX \bv$.	
}, i.e., they possess frame-independent meaning purely attached to the differential structure of the spacetime. 
 The torsion 2-form is then abstractly defined as 	
\beq \label{Eq:GeoHiggs}
 	\bcT \equiv \tilde{\Del} - \bcL
\eeq where the tilde in $\tilde{\Del}$ signifies alternation; i.e., the antisymmetrized feed of arguments to the $\Del$ operator:
 $\tilde{\Del}_{\bX}\,\bY \equiv \Del_{\bX}\,\bY - \Del_{\bY}\,\bX$.

	With {\bf e} and $\Del$ as basic building blocks, 
 the genius of Elie Cartan conceived the notion of vector-valued differential p-forms: mixed tensors, once-contravariant (the ``vector part"), 
 p-covariant tensors (the p-form part). 
 Applied to vierbeins as frame ({\it vector}) fields, this machinery bears fruit to the abstract notion of {\it solder forms}
 as a way to match the {\it intrinsic} tangent space of a spacetime to the {\it trivialization of such} as a fibered product
 of appropriate (usually Euclidean) trivializations of the space.
 With matrix multiplication implied, explicitly we have: 
\beq	\label{Eq:Build_Curv}
	\Del \be = \be \, \omega ~~~{\rm and}~~~ \Del \, \Del \be = \be (\omega \wedge \omega + d \omega) = \be \, \cS,
\eeq		
 where the vierbein, $\be$, are vector fields;
 the connection 1-forms, $\omega$, are matrix-valued 1-forms 
 (in fact, elements of the {\it dual} vector space ; i.e. linear functionals on the vector space with values on the real numbers)
 and where the curvature 2-form, $\cS$ is matrix-valued 
 (note that {\bf boldface} typesetting here and below implies either vectors or vector fields whereas standard typesetting may
 imply either forms valued on vector fields or their duals: matrices).
 In practice, $\cS$ may be interpreted as a matrix with entries as 2-forms instead of the other way around (see, e.g., Frankel\cite{Frankel97})

	On the other hand, the notion of Lie derivative in combination with 						
 the machinery of exterior calculus gives rise to the following remarkable theorem:
 Let \bX~ and \bY~ be vectors (not necessarily fields) at a point $p$ and let $\alpha^1$ be any 1-form.  If the vectors can be extended to a 
 neighborhood of $p$ in a smooth manner--e.g., possibly but not necessarily as vector {\it fields}--we then have:
\[ 	d \alpha^1 (\bX, \bY) = \bcL_\bX(\alpha^1(\bY)) - \bcL_\bY(\alpha^1(\bX)) - \alpha^1( \bcL_\bX (\bY)).
\]
 With $\bvarep$ and $\bcL$ as basic building blocks, this theorem yields 
 (note that when applied to a basis of orthogonal ``holonomic" frames,  $\varepsilon$ are simply the inverse vierbein):
\bea	
\label{Eq:1st-Struc_Eq}
d \varepsilon^i (\be_j,\be_k) 	&=&  \bcL_{\be_j}(\varepsilon^i(\be_k)) - \bcL_{\be_k}(\varepsilon^i(\be_j)) - \varepsilon^i (\bcL_{e_j} e_k) \cr
					&&\cr
					&=&  - \varepsilon^i \left( \left\{\tilde{\Del} - \bcT \right\}_{e_j} e_k \right) \cr
					&&\cr
	d \varepsilon 			&=& - \omega \wedge \varepsilon + \cT 	
\eea
 This is the content of the Cartan's celebrated structure equation.

 	In summary, the machinery of exterior calculus in the language of differential forms allows for two natural differential structures to be bestowed
 upon a spacetime by virtue of the covariant derivative and the Lie derivative.  The derived set of structure equations embodied by
 Eqs[\ref{Eq:Build_Curv} \& \ref{Eq:1st-Struc_Eq}]:
\[
 \cT = d \varepsilon + \omega \wedge \varepsilon ~~~{\rm and}~~~ \cS = d\omega + \omega \wedge \omega
\]
 follow from the canonical application of each of these operators while torsion represents the {\it difference} in differential structure between the two.

	We now cast a spotlight on the origin of the ``internal space" within the context of the Palatini formulation where 
 the vierbein's flat indices (and those of its covariant derivatives in the sense of Eq[\ref{Eq:Build_Curv}])
 identified such a space (or its dual in the sense of linear functionals on the space) with values that the 1-form takes upon.  	
 Notably, these definitions do not make (implicit nor explicit) use of the Lie derivative! 
 Yet, it is the Lie derivative that brings about the notion of torsion into the geometrical picture. 
 This brief analysis seems to indicate that there is no room in the fibered product construction to accommodate for torsion.
 Still, a compelling  desire to understand Einstein's equation at an algebraic level motivates an urgent need to incorporate the torsion into the picture.
 We hope that this note is a small step in the right direction.

{\footnotesize
\begin{flushright} {\it --$\cN\!$umen $\ae$theris $\ae$ternum {\cI}ntersunt--} \end{flushright}
} 

\section{Some brief concluding remarks}

	One of the deepest mysteries of the Universe is embodied by the so-called dark energy problem.
 It seems almost funny that Albert Einstein inadvertently discovered such a degree of freedom in his profound and celebrated equation.
 Granted the tremendous success of general relativity in predicting direct, testable gravitational phenomena at moderate fields,
 e.g., the precession of Mercury's perihelium, the gravitational wave looses in binary pulsar PSR J0737-3039; 
one feels naturally obliged to conform with the standard form of Einstein's equation!
 Yet, we have emphasized throughout this note that maintaining balance in a single equation necessarily implies mixing the Lie algebra and the abelian space 
 in some fashion. 

  	Commencing under the premise that to do physics one needs rulers and clocks, 
 the Lorentz group is handed to us by God as the fundamental symmetry of spacetime.		
 It seems only fitting that the erection of curved spacetimes be contingent upon rigorous adherence to its ample geometrical structure.
 Our thesis sustains that implementing such a symmetry locally--in the sense of diffeomorphism invariance--amounts
 to ``choosing" an appropriate algebro-geometric--differential tool as to make algebraic sense of Einstein's equation or of a sensible evolution of such 
 an equation.  In this regard,
 the theorist dream is to find a (renormalizable) gravitational action that incorporates torsion, a (geometric) Higgs field and which naturally gives rise 
 to a cosmological term (or a sensible evolution of such a term) upon variation of the dynamical fields. 

 	The fundamental geometric object in this business is the vierbein as a frame field ({\it not} its dual) and the standard differential tool is
 the covariant derivative.
 The Lie algebra valued 1-form of spin connection and its derived 2-form of curvature are readily built from these two pieces while referring to an internal space 
 with ``global" Lorentz structure. 
 In this internal space picture, the torsion 2-form, which is derived from the Lie derivative as an equally important differential tool,
 is similarly valued on the Lie algebra of $\bar{\cL}$.
 Yet, its inclusion into the geometrical framework precludes the employment of a {\it principal frame bundle} to fit coordinate patch expressions
 for the tangent bundle ``nicely" as a trivially fibered product.  
 On the other hand,
 since Tor-functors measure the degree to which the tensor product of a left module with a short exact sequence of right modules fails itself to be 
exact\cite{Car&Eil56}, inclusion of such homological considerations may be required to make sense of the tangent bundle as a fibered product.

 	In addition to these two intrinsically differential tools, we have also discussed two {\it algebraic} derivation tools that could be important in the 
 quantization of the theory:
 i- Fox derivatives within the framework of formal representation theory and ii- left-$\alpha$ derivations from non-commutative ring theory.
 Either of these could play an equally important role in the interpretation of Einstein's equation as a means to a dynamical iteration of 
 background-independent, curved spacetimes.  
 At this point, it is not clear which set of operators will lead to the sought after explication of Einstein's equation
 but we can conjecture two possibilities in accordance to whether one prefers to stay within the ``standard" group theoretic framework
 or fully depart into the realm of non-commutative ring theory.
 Under the former scenario, one promotes the Fox derivatives in the (Lorentz) group (in lieu of covariant derivatives) 
 to full derivations in the group ring and tries to make peace with the algebraic structure of Einstein's equation. 
 On the other hand, under the fully ring theoretic scenario, one does the same with the left-$\alpha$ derivations while accommodating for the possibility that
 spacetime as a vector space could become a non-commutative bimodule.
 
 	It seems rather promising that existing literature with the framework of a Cartan connection hints at the formal coupling of spin to torsion
 at the classical level. 
 Indeed, the link in the context of the semidirect product has been closely examined in particular by Sternberg and 
collaborators\cite{Stern&Ungar79, Rapo&Ster84, Sternb85}.
 These authors conclude that the coupling of spin to torsion may be attributed to the null foliation of the pre-symplectic structure 
 (allowing for degeneracy in the symplectic form) laying in the midst of the group and the space.
 In their work, this is parametrized by a (poorly understood) $\bar{\cL}$-invariant condition on the suborbits of the foliation. 
 In the present work, we formally motivate a curved version of Poincar\'e space as a semidirect product: a non-trivial group extension of 
 $\bar{\cL}$ by its affine derivation space.
 Close scrutiny of the invariance condition under the present framework seems fertile ground for future research. 

	Lastly, recall that conforming with the canonical formulation of general relativity {\it via} Noether's theorem leads to an asymmetric stress-energy 
 tensor which, in flat spacetime, may be symmetrized by restoring spin current conservation as 
Belinfante and Rosenfeld\cite{Belin40, Rosen40}
have shown.  Although in either the purely quantum or the purely relativistic regimes both the spin currents and the torsion are expected to be negligible,  
 we conjecture that the fine balance embodied by the cosmological constant term in the action is lost in the Quantum Gravity regime where the spin current
 and torsion terms may not be negligible. 
 Indeed, this lost of balance may drive inflation under a high energy phase transition at the very early epochs near the Plank scale.


\end{document}